\documentclass{article}

\usepackage[preprint]{neurips_2026}

\usepackage[utf8]{inputenc} 
\usepackage[T1]{fontenc}    
\usepackage{hyperref}       
\usepackage{url}            
\usepackage{booktabs}       
\usepackage{amsfonts}       
\usepackage{nicefrac}       
\usepackage{microtype}      
\usepackage[dvipsnames]{xcolor}
\DeclareUnicodeCharacter{4E01}{}
\DeclareUnicodeCharacter{6E90}{}
\DeclareUnicodeCharacter{68EE}{}
\usepackage{tikz}         
\usetikzlibrary{arrows.meta, calc}
\usepackage{graphicx}      
\usepackage{wrapfig}        
\usepackage{bm}

\hypersetup{
    colorlinks = true,
    linkbordercolor = white,
    citecolor = MidnightBlue,
    linkcolor = MidnightBlue,
    urlcolor = OliveGreen,
}


\title{\textsc{Mushy}: Multimodal Flow-Based Amortized Bayesian Inference for Spectroscopic Data Fusion}

\author{%
  Jeff Shen \\
  Princeton University\\
  Princeton, NJ 08540 \\
  \texttt{shenjeff@princeton.edu} \\
}

\begin{document}

\maketitle

\begin{abstract}
  We are in an era of unprecedented spectroscopic data availability, with surveys such as APOGEE, GALAH, and DESI observing millions of stars.
  However, data are heterogeneous and to date, efforts to exploit synergies between surveys have been limited.
  In particular, there is a lack of principled methods that combine spectroscopic data from different surveys to improve inference of physical parameters.
  In this work we present a new method, \textsc{Mushy}, for spectroscopic data fusion.
  \textsc{Mushy} is an amortized Bayesian inference method, using a rectified flow model with multimodal inputs to learn a posterior distribution over physical parameters.
  Using synthetic spectra generated via ATLAS12/SYNTHE, including realistic instrumental effects and noise, we demonstrate that \textsc{Mushy} can precisely infer physical parameters with well-calibrated posteriors from any combination of surveys, with multimodal inputs providing stronger constraints on the posterior than single-survey data alone.
  This work paves the way for principled combination of spectroscopic data from heterogeneous sources.
\end{abstract}

\section{Introduction}
\label{sec:intro}

How do we make the most of the data that we have?
We have a wealth of data from the huge number of spectroscopic surveys that are currently online \citep[e.g.,][]{Majewski2017, Buder2021, DESICollaboration2016}, and the amount of heterogeneous data available is only growing as more telescopes come online \citep[e.g.,][]{Kollmeier2017}. 
We would like to infer physical parameters from these data in order to gain more insight into, for example, the formation and evolution of the Milky Way \citep{Helmi2018, Rix2022}.
This is typically done by survey-specific pipelines, e.g., ASPCAP for APOGEE \citep{GarciaPerez2016}, independently for each survey.
However, surveys often observe the same stars and provide complementary data, so siloed inference is suboptimal.
We would like to combine data from different surveys to improve our inference.

Here we present the \textsc{Mushy} model for doing so.
Broadly, \textsc{Mushy} follows the amortized Bayesian inference paradigm, using a rectified flow model with multimodal conditioning to learn a posterior distribution over physical parameters given spectra from varying combinations of surveys.
At inference time, \textsc{Mushy} can ingest any combination of surveys and rapidly produce a posterior distribution that is both well-calibrated and takes advantage of the complementary information from the different surveys.
To our knowledge, this is the first work to do principled multimodal inference for spectroscopic data fusion.

\paragraph{Related work}
Multimodal models are increasingly used in astrophysics to combine heterogeneous data into shared representations \citep{Parker_2024, parker2025aion, Shen2025, Shen2026}; we build on one of them, the spectral tokenizer of \cite{Shen2025}, as our encoder.
To turn such representations into physical parameters, we use simulation-based inference (SBI), which enables Bayesian inference when the likelihood is intractable but data can be simulated from the parameters \citep{papamakarios2018sbi, Cranmer_2020, lueckmann2021}.
Specifically, we perform amortized posterior estimation: a conditional generative model is trained on simulated pairs to approximate the posterior, so that inference on new data costs a single forward pass \citep[e.g.,][]{radev2020}.
Spectral \citep{Ting2019, rozanski2024} and atmospheric \citep{li2025a1} emulators approach the problem from the other direction: they replace the expensive simulator with a fast neural approximation, making direct fitting with MCMC or optimization more tractable, but still require a (possibly still expensive) fit for every star.
Amortizing inference removes this per-star cost.

\section{Data}
\label{sec:data}

Our basic setup is that we have a single underlying physical model, parameterized by $\theta$, that generates the data, and we have multiple surveys that provide different views of the data $x_1, \ldots, x_n$.
These multiple incomplete views can be generated via a simulator (physics model) $S: \theta \mapsto x_1, \ldots, x_n$, which we assume to be a good approximation of reality (despite the inevitable sim-to-real gap, i.e., imperfectly specified physics).

Synthetic spectra are generated with ATLAS12 and SYNTHE via \texttt{pykurucz} \citep{kurucz2005, kim2026pykurucz}.
ATLAS12 is responsible for generating the stellar atmosphere model from physical parameters, and SYNTHE generates the synthetic spectrum from the stellar atmosphere model.
Here we select $\theta$ to be 21-dimensional: $T_{\rm eff}$, $\log g$, $v_{\rm micro}$, [Fe/H], and 17 [X/Fe] (C, N, O, Na, Mg, Al, Si, S, K, Ca, Ti, V, Cr, Mn, Co, Ni, Ce).
ATLAS12/SYNTHE gives us a fully self-consistent and ab-initio simulator, eliminating the need for the typical interpolation on pre-computed atmosphere grids, or emulators for spectral synthesis, both of which can introduce additional errors.

We generate synthetic spectra for three surveys: APOGEE, GALAH, and DESI.
These surveys provide complementary information about $\theta$;
for example, APOGEE is in the H-band and is sensitive to C, N, and $\alpha$ elements, while GALAH is in the optical and is sensitive to Na, V, and Cr.
For each $\theta$, we generate a synthetic spectrum that covers the wavelength range of all three surveys (358--985\,nm and 1500--1710\,nm) at a high resolution ($R=100{,}000$) and resample it to the native wavelength grids of each survey.
We use APOGEE DR17 line lists from 1500--1700\,nm \citep{Smith2021}, Gaia-ESO line lists \citep{Heiter2021} from 475--690\,nm and 840--895\,nm, and Kurucz \texttt{gfall}\footnote{\protect{\url{http://kurucz.harvard.edu/linelists.html}}} elsewhere. 
We simulate instrumental effects and noise to generate synthetic spectra that resemble real survey data via \textit{donor injection}.
Each synthetic spectrum is injected into a real ``donor'' spectrum from the survey, which provides the instrument response (e.g., line-spread function), continuum shape, noise level, and bad-pixel mask. 
We repeat this process for 50 realizations per survey per $\theta$.

The prior over $\theta$ is a mixture of a stellar isochrone population, a uniform distribution over the parameter box, a small fraction of rare populations (carbon-enhanced, globular-cluster, r-process), as well as a small fraction of $\theta$ at the pipeline parameters of real survey stars.
We generate a training set of ${\sim}$48K $\theta$ with all three surveys, and hold out ${\sim}1\%$ for validation.

\section{Method}
\label{sec:method}

\begin{figure}[t]
  \centering
  \resizebox{\linewidth}{!}{\definecolor{apogee}{HTML}{D55E00}
\definecolor{galah}{HTML}{009E73}
\definecolor{desi}{HTML}{0072B2}
\definecolor{frozen}{HTML}{56B4E9}
\begin{tikzpicture}[
  font=\small,
  x=1.3cm, 
  >={Stealth[length=4.5pt, width=3.6pt]},
  arrow/.style={->, line width=0.6pt, draw=black!65},
  block/.style={rectangle, rounded corners=3pt, draw=black!55, line width=0.6pt,
                fill=black!4, align=center, minimum height=1.15cm,
                minimum width=1.85cm, inner sep=3pt},
  note/.style={font=\scriptsize, text=black!60, align=center},
  spec/.style={line width=0.5pt, line join=round, line cap=round},
]

\input{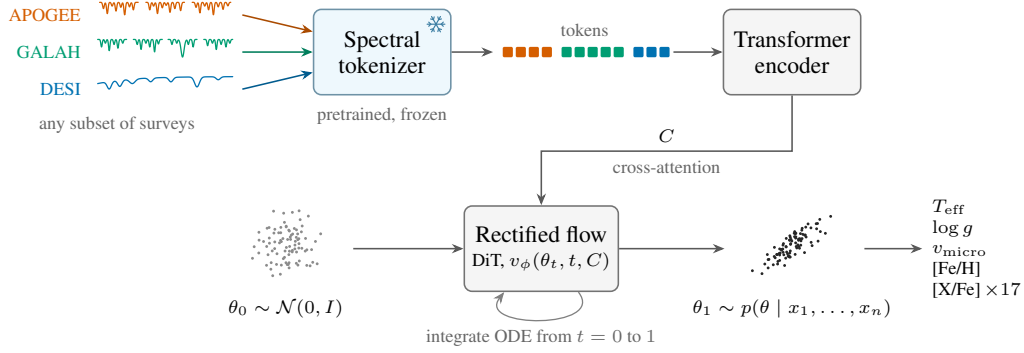}
\node[anchor=east, font=\scriptsize, text=apogee] at (0.92, 0.50) {APOGEE};
\node[anchor=east, font=\scriptsize, text=galah]  at (0.92, 0.00) {GALAH};
\node[anchor=east, font=\scriptsize, text=desi]   at (0.92,-0.50) {DESI};
\node[note] at (1.2,-0.98) {any subset of surveys};

\node[block, fill=frozen!10, draw=frozen!55!black!70] (tok) at (3.95,0)
  {Spectral\\tokenizer};
\coordinate (sf) at ($(tok.north east)+(-0.21cm,-0.21cm)$);
\foreach \a in {0,60,...,300}{
  \draw[frozen!70!black, line width=0.45pt, line cap=round] (sf) -- ++(\a:0.12cm);
  \draw[frozen!70!black, line width=0.45pt, line cap=round]
    ($(sf)+(\a:0.07cm)$) -- ++(\a+45:0.04cm);
  \draw[frozen!70!black, line width=0.45pt, line cap=round]
    ($(sf)+(\a:0.07cm)$) -- ++(\a-45:0.04cm);
}
\node[note, below=1pt] at (tok.south) {pretrained, frozen};
\foreach \y/\c in {0.5/apogee, 0/galah, -0.5/desi}
  \draw[arrow, draw=\c!80!black] (2.5,\y) -- ($(tok.west)+(0,0.55*\y)$);

\foreach \k/\c/\grp in {0/apogee/0, 1/apogee/0, 2/apogee/0, 3/apogee/0,
                        4/galah/1, 5/galah/1, 6/galah/1, 7/galah/1, 8/galah/1,
                        9/desi/2, 10/desi/2, 11/desi/2}{
  \pgfmathsetmacro{\xk}{5.2 + 0.135*\k + 0.07*\grp}
  \fill[\c, rounded corners=0.6pt] (\xk,-0.07) rectangle ++(0.105,0.14);
}
\node[note] at (6.05,0.3) {tokens};
\draw[arrow] (tok.east) -- (5.1,0);

\node[block] (enc) at (8.2,0) {Transformer\\encoder};
\draw[arrow] (6.97,0) -- (enc.west); 

\node[block, minimum width=2.0cm] (flow) at (5.6,-2.65)
  {Rectified flow\\[-1pt] {\scriptsize DiT,\ $v_\phi(\theta_t, t, C)$}};
\draw[arrow] (enc.south) -- ++(0,-0.75) -| (flow.north)
  node[pos=0.25, above, font=\scriptsize]{$C$}
  node[pos=0.25, below, note]{cross-attention};
\draw[->, line width=0.5pt, draw=black!50]
  ([xshift=5mm]flow.south) .. controls +(0.3,-0.5) and +(-0.3,-0.5)
  .. ([xshift=-5mm]flow.south);
\node[note] at ($(flow.south)+(0,-0.62)$) {integrate ODE from $t=0$ to $1$};

\coordinate (noise) at (2.95,-2.6);
\pgfmathsetseed{7}
\foreach \i in {1,...,90}{
  \pgfmathsetmacro{\ra}{min(sqrt(-2*ln(0.001 + 0.998*rnd)), 2.6)}
  \pgfmathsetmacro{\an}{360*rnd}
  \fill[black!45] ($(noise)+(\an:{0.19cm*\ra})$)
    circle[radius=0.55pt];
}
\node[font=\scriptsize] at (2.95,-3.45) {$\theta_0 \sim \mathcal{N}(0, I)$};
\draw[arrow] (3.65,-2.65) -- (flow.west);

\coordinate (post) at (8.2,-2.6);
\pgfmathsetseed{11}
\foreach \i in {1,...,90}{
  \pgfmathsetmacro{\ra}{min(sqrt(-2*ln(0.001 + 0.998*rnd)), 2.6)}
  \pgfmathsetmacro{\an}{360*rnd}
  \pgfmathsetmacro{\za}{\ra*cos(\an)}
  \pgfmathsetmacro{\zb}{\ra*sin(\an)}
  \fill[black!85] ($(post)+(35:{0.25cm*\za})+(125:{0.07cm*\zb})$)
    circle[radius=0.55pt];
}
\node[font=\scriptsize] at (8.2,-3.45) {$\theta_1 \sim p(\theta \mid x_1, \ldots, x_n)$};
\draw[arrow] (flow.east) -- (7.5,-2.65);

\draw[arrow] (8.95,-2.65) -- (9.55,-2.65);
\node[anchor=west, align=left, font=\scriptsize, inner sep=1pt] at (9.62,-2.65)
  {$T_{\rm eff}$\\ $\log g$\\ $v_{\rm micro}$\\ {[Fe/H]}\\ {[X/Fe]}\,$\times 17$};

\end{tikzpicture}}
  \caption{The \textsc{Mushy} architecture. Spectra from any subset of surveys are encoded by a frozen, pretrained spectral tokenizer; the tokens are concatenated and passed through a transformer encoder to give the conditioning signal $C$. A conditional rectified flow with a DiT backbone predicts the velocity $v_\phi(\theta_t, t, C)$, attending to $C$ through cross-attention. Integrating the ODE from $t=0$ to $1$ transports Gaussian noise $\theta_0$ to samples from the posterior over the 21 stellar parameters.}
  \label{fig:architecture}
  \vspace{-0.25cm}
\end{figure}

We adopt an amortized Bayesian inference approach \citep[e.g.,][]{Cranmer_2020}.
Given our simulator $S: \theta \mapsto x_1, \ldots, x_n$ that maps physical parameters $\theta$ to multimodal data $x_1, \ldots, x_n$, we would like to learn a posterior distribution $p(\theta \mid x_1, \ldots, x_n)$ over physical parameters given data.
Concretely, \textsc{Mushy} is a conditional rectified flow \citep{Liu2022a} with a DiT backbone \citep{Peebles2023}.
Figure \ref{fig:architecture} shows the architecture of \textsc{Mushy}, and here we briefly discuss the components and design choices.

A key question is: how do we combine the data from different surveys ($x_1, \ldots, x_n$) and perform inference with any combination/subset of surveys?
There are two issues to consider: (1) the data from different surveys are heterogeneous, and (2) our conditioning method needs to be flexible to the number of surveys present.

For the first, we use the universal spectral tokenizer from \cite{Shen2025}.
This model is pretrained in a self-supervised manner on real spectral data (including APOGEE, DESI, and GALAH) and is designed to convert the heterogeneous input spectra into a common latent space, which has been shown to carry useful information about physical parameters;
we can then combine the latent representations by simple concatenation.
We use a lightweight transformer encoder on top of the conditioning inputs to model interactions between the tokens from the different surveys.

The second issue we tackle with cross-attention in the DiT backbone, which allows us to condition on arbitrary length inputs.
The tokens of the observed surveys $\mathcal{A}$ are embedded, concatenated and encoded into a conditioning sequence $C \in \mathbb{R}^{L \times d}$, $L = \sum_{s \in \mathcal{A}} L_s$. The noisy parameters $\theta_t$ enter the DiT as a single token $u$, which in every block cross-attends to $C$, with $t$ entering through adaptive RMS norm.
During training, we use modality dropout to randomly drop a subset of surveys for each example, so that the conditioning signal given to the model varies and the model learns to handle any combination of surveys.

\begin{wraptable}{r}{0.515\linewidth}
  \setlength{\belowcaptionskip}{5pt}
  \vspace{-10pt}
  \centering
  \caption{$\sigma_{\rm MAD}$ of the posterior median on held-out data. Subscripts show bootstrap standard errors. Full table in Table~\ref{tab:sigma_mad_all}.}
\label{tab:sigma_mad_main}
{\footnotesize
\setlength{\tabcolsep}{3pt}
\begin{tabular}{lllll}
    \toprule
    Surveys & $T_{\rm eff}$ (K) & $\log g$ & [Fe/H] & [Mg/Fe] \\
    \midrule
    APOGEE & 67\textsubscript{$\pm$3} & 0.075\textsubscript{$\pm$0.003} & 0.043\textsubscript{$\pm$0.002} & 0.038\textsubscript{$\pm$0.002} \\
    GALAH & 49\textsubscript{$\pm$2} & 0.155\textsubscript{$\pm$0.007} & 0.069\textsubscript{$\pm$0.003} & 0.079\textsubscript{$\pm$0.004} \\
    DESI & 72\textsubscript{$\pm$4} & 0.188\textsubscript{$\pm$0.009} & 0.099\textsubscript{$\pm$0.005} & 0.094\textsubscript{$\pm$0.005} \\
    All three & \textbf{36}\textsubscript{$\pm$2} & \textbf{0.063}\textsubscript{$\pm$0.002} & \textbf{0.034}\textsubscript{$\pm$0.002} & \textbf{0.036}\textsubscript{$\pm$0.002} \\
    \bottomrule
\end{tabular}}

\end{wraptable}
We keep the pretrained spectral tokenizer frozen and train a small rectified flow on top of it.
Our model is lightweight: it consists of just 2 conditioner blocks and 2 DiT blocks, each with an embedding dimension of 384 and 8 attention heads, for a total of 12M trainable parameters.
It trains in less than 30 minutes on one $8\times$H200 node using a standard $\bm{v}$-pred parameterization and $\bm{v}$-loss objective \citep[e.g.,][]{li2026basics}.
We use an AdamW optimizer with $(\beta_1, \beta_2) = (0.9, 0.95)$, weight decay of $0.1$, and a warmup-stable-decay learning rate schedule set to 10\% linear warmup and 10\% linear decay over a total of 20 epochs, with a peak learning rate of $3\times10^{-4}$ \citep{Loshchilov2019, hu2024wsd}.

\section{Results}
\label{sec:results}

When we apply \textsc{Mushy} to held-out simulation stars with known $\theta$, we can compare the inferred posterior distribution with the true $\theta$.
We quantify the accuracy of the inference by computing $\sigma_{\rm MAD} = 1.4826 \times {\rm median}(|\hat{\theta} - \theta_{\rm true}|)$, a robust measure of dispersion where $\hat{\theta}$ is taken to be the posterior median and the $1.4826$ factor makes $\sigma_{\rm MAD}$ comparable to the standard deviation for a Gaussian distribution.

We show in Table \ref{tab:sigma_mad_main} that \textsc{Mushy} can precisely infer physical parameters with both single-survey inputs and multi-survey inputs, and that multi-survey inference always provides stronger constraints on the posterior distribution than single-survey inference.
In particular, each constituent survey provides different constraints on different physical parameters---for example, $T_{\rm eff}$ is best constrained by GALAH, while $\log g$ is best constrained by APOGEE---and combining them allows us to take the best of each survey and improve our inference.
In Table \ref{tab:sigma_mad_all} in Appendix \ref{app:sigma_mad}, we show the $\sigma_{\rm MAD}$ for all 21 parameters.
We also show in Figure \ref{fig:calibration} that the posteriors are generally well-calibrated, with the exception of a few elements (K, Ni, Mn) that are mildly overconfident.

\begin{wrapfigure}[23]{r}{0.53\linewidth}
  \vspace{-15pt} 
  \centering
  \includegraphics[width=\linewidth]{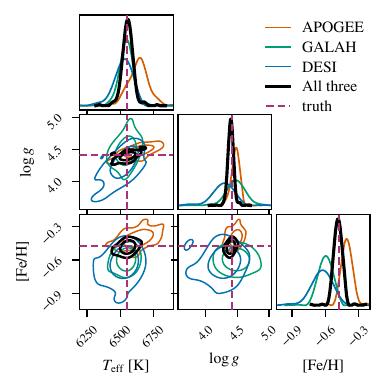}
  \caption{Posteriors for one held-out star under each survey alone and all three together. Contours enclose 68\% and 95\%; dashed lines mark the truth.}
  \label{fig:pairs_heldout}
\end{wrapfigure}
\begin{figure}[h]
  \centering
  \includegraphics[width=\linewidth]{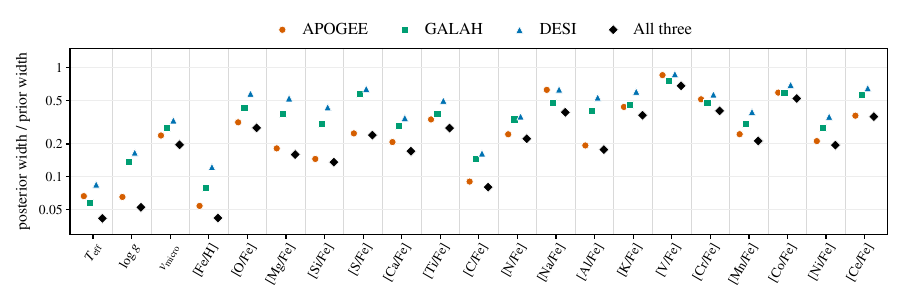}
  \caption{Information gain per parameter on held-out data; lower is better. The points show the median posterior width over the prior width (information gain), for each survey alone and for all three together. A value of 1 means the spectra add nothing to the prior. Multimodal inference always provides more information than single-survey inference.}
  \label{fig:infogain}
  \vspace{-0.25cm}
\end{figure}

We show in Figure \ref{fig:pairs_heldout} the posteriors for key atmospheric parameters for a single held-out star, as estimated under each survey alone and all three together; Figure~\ref{fig:pairs_heldout_all} shows all 21 parameters for the same star.
We again find that multimodal inference outperforms single-survey inference: we obtain tighter posteriors that are centered on the truth when we combine all three surveys, while single-survey posteriors are broader and offset from the truth.
We can quantify this information gain of a given posterior relative to the prior, calculated as the ratio of the posterior width to the prior width.
We do this in Figure \ref{fig:infogain} and find that GALAH gives the narrowest posteriors for $T_{\rm eff}$, Na, V and Cr, and APOGEE for $\log g$, [Fe/H], C, N and most other elements; all three together give the narrowest posterior for every parameter.
Broadly, the multi-survey posteriors are narrower than the single-survey posteriors, which are in turn narrower than the prior, indicating that including more surveys provides more information about the physical parameters.

\section{Conclusion}
\label{sec:conclusion}

We have presented \textsc{Mushy}, which to our knowledge is the first model to do principled multimodal inference for spectroscopic data fusion.
Built on an amortized Bayesian inference scheme and stellar atmosphere and spectral synthesis simulators, \textsc{Mushy} is a single network that can precisely infer physical parameters with well-calibrated posteriors from arbitrary combinations of surveys and can take advantage of the complementary information from the different surveys to improve inference.
Further, we can quantify the information gain from each survey to understand what each survey contributes to the inference of each physical parameter.
Tools like \textsc{Mushy} will be increasingly important as the volume of heterogeneous data continues to grow.

\paragraph{Limitations and future work}

Standard SBI limitations apply: the posteriors are only as good as the simulator, and the model has no support for stars outside the simulator's prior.
There is a sim-to-real gap due to missing/misspecified physics in the simulator (e.g., imperfect line list), and real spectra are distinguishable from simulated spectra (a real vs simulated classifier achieves greater-than-chance accuracy).
The simulator's residuals could perhaps be learned from real spectra and be applied as a correction to the simulator.
Further, applying the model to real spectra would be the obvious and most important next step, with validation of posteriors against reliable external estimates (e.g., asteroseismic $\log g$ and open-cluster metallicities).

\begin{ack}
J.S. thanks Lindsay Smith for helpful comments on the manuscript.
J.S. is supported by the Natural Sciences and Engineering Research Council of Canada (NSERC), funding reference number 587652. 
In addition, the computations in this work were carried out on, and the data hosted on equipment supported by the Scientific Computing Core at the Flatiron Institute, a division of the Simons Foundation. 
\end{ack}

\bibliographystyle{unsrtnat}
\bibliography{../references, ../references2}

\appendix

\renewcommand{\topfraction}{0.95}
\renewcommand{\bottomfraction}{0.95}
\renewcommand{\textfraction}{0.05}
\renewcommand{\floatpagefraction}{0.8}
\setcounter{topnumber}{3}
\setcounter{totalnumber}{4}

\clearpage
\section{Recovery of all parameters}
\label{app:sigma_mad}

\begin{table}[ht]
  \centering
  \caption{Recovery for all 21 parameters and all seven survey combinations: $\sigma_{\rm MAD}$ of the posterior median against the truth, on 882 held-out $\theta$. Lower is better. A+G is APOGEE with GALAH, A+D APOGEE with DESI, G+D GALAH with DESI. $T_{\rm eff}$ is in K, $v_{\rm micro}$ in km/s, the rest in dex. Subscripts are bootstrap standard errors over the held-out stars. In each row the lowest value is in bold, and entries statistically indistinguishable from it (95\% paired bootstrap) are underlined.}
  \label{tab:sigma_mad_all}
  \footnotesize
  \setlength{\tabcolsep}{3.5pt}
  \begin{tabular}{lrrrrrrr}
    \toprule
    & \multicolumn{3}{c}{Single survey} & \multicolumn{3}{c}{Pair} & \\
    \cmidrule(lr){2-4} \cmidrule(lr){5-7}
    Parameter & APOGEE & GALAH & DESI & A+G & A+D & G+D & All \\
    \midrule
    {$T_{\rm eff}$ (K)} & 67\textsubscript{$\pm$3} & 49\textsubscript{$\pm$2} & 72\textsubscript{$\pm$4} & 43\textsubscript{$\pm$2} & 48\textsubscript{$\pm$2} & 43\textsubscript{$\pm$2} & \textbf{36}\textsubscript{$\pm$2} \\
    {$\log g$} & 0.075\textsubscript{$\pm$0.003} & 0.155\textsubscript{$\pm$0.007} & 0.188\textsubscript{$\pm$0.009} & \underline{0.067}\textsubscript{$\pm$0.003} & \underline{0.067}\textsubscript{$\pm$0.003} & 0.134\textsubscript{$\pm$0.006} & \textbf{0.063}\textsubscript{$\pm$0.002} \\
    {$v_{\rm micro}$ (km/s)} & 0.172\textsubscript{$\pm$0.008} & 0.163\textsubscript{$\pm$0.008} & 0.208\textsubscript{$\pm$0.010} & \underline{0.142}\textsubscript{$\pm$0.006} & 0.160\textsubscript{$\pm$0.007} & 0.159\textsubscript{$\pm$0.008} & \textbf{0.140}\textsubscript{$\pm$0.006} \\
    {[Fe/H]} & 0.043\textsubscript{$\pm$0.002} & 0.069\textsubscript{$\pm$0.003} & 0.099\textsubscript{$\pm$0.005} & \underline{0.035}\textsubscript{$\pm$0.002} & 0.040\textsubscript{$\pm$0.002} & 0.061\textsubscript{$\pm$0.003} & \textbf{0.034}\textsubscript{$\pm$0.002} \\
    \addlinespace[3pt]
    {[O/Fe]} & 0.072\textsubscript{$\pm$0.005} & 0.095\textsubscript{$\pm$0.006} & 0.116\textsubscript{$\pm$0.005} & \underline{0.066}\textsubscript{$\pm$0.004} & 0.073\textsubscript{$\pm$0.005} & 0.090\textsubscript{$\pm$0.005} & \textbf{0.065}\textsubscript{$\pm$0.004} \\
    {[Mg/Fe]} & \underline{0.038}\textsubscript{$\pm$0.002} & 0.079\textsubscript{$\pm$0.004} & 0.094\textsubscript{$\pm$0.005} & \textbf{0.036}\textsubscript{$\pm$0.002} & \underline{0.037}\textsubscript{$\pm$0.002} & 0.068\textsubscript{$\pm$0.003} & \textbf{0.036}\textsubscript{$\pm$0.002} \\
    {[Si/Fe]} & \underline{0.032}\textsubscript{$\pm$0.001} & 0.059\textsubscript{$\pm$0.003} & 0.073\textsubscript{$\pm$0.004} & \underline{0.032}\textsubscript{$\pm$0.002} & \underline{0.031}\textsubscript{$\pm$0.002} & 0.060\textsubscript{$\pm$0.003} & \textbf{0.030}\textsubscript{$\pm$0.002} \\
    {[S/Fe]} & 0.075\textsubscript{$\pm$0.004} & 0.141\textsubscript{$\pm$0.008} & 0.153\textsubscript{$\pm$0.008} & \textbf{0.066}\textsubscript{$\pm$0.003} & \underline{0.069}\textsubscript{$\pm$0.004} & 0.134\textsubscript{$\pm$0.007} & \textbf{0.066}\textsubscript{$\pm$0.003} \\
    {[Ca/Fe]} & 0.053\textsubscript{$\pm$0.004} & 0.063\textsubscript{$\pm$0.003} & 0.063\textsubscript{$\pm$0.003} & 0.049\textsubscript{$\pm$0.002} & \underline{0.044}\textsubscript{$\pm$0.002} & 0.051\textsubscript{$\pm$0.003} & \textbf{0.042}\textsubscript{$\pm$0.002} \\
    {[Ti/Fe]} & 0.080\textsubscript{$\pm$0.004} & 0.086\textsubscript{$\pm$0.004} & 0.105\textsubscript{$\pm$0.005} & 0.078\textsubscript{$\pm$0.004} & 0.078\textsubscript{$\pm$0.005} & 0.078\textsubscript{$\pm$0.003} & \textbf{0.070}\textsubscript{$\pm$0.004} \\
    \addlinespace[3pt]
    {[C/Fe]} & 0.063\textsubscript{$\pm$0.003} & 0.104\textsubscript{$\pm$0.005} & 0.108\textsubscript{$\pm$0.006} & \underline{0.062}\textsubscript{$\pm$0.003} & \underline{0.061}\textsubscript{$\pm$0.003} & 0.088\textsubscript{$\pm$0.005} & \textbf{0.056}\textsubscript{$\pm$0.003} \\
    {[N/Fe]} & 0.120\textsubscript{$\pm$0.004} & 0.145\textsubscript{$\pm$0.007} & 0.135\textsubscript{$\pm$0.006} & \underline{0.110}\textsubscript{$\pm$0.005} & 0.113\textsubscript{$\pm$0.004} & 0.130\textsubscript{$\pm$0.006} & \textbf{0.106}\textsubscript{$\pm$0.005} \\
    \addlinespace[3pt]
    {[Na/Fe]} & 0.188\textsubscript{$\pm$0.010} & 0.135\textsubscript{$\pm$0.006} & 0.167\textsubscript{$\pm$0.007} & 0.129\textsubscript{$\pm$0.006} & 0.155\textsubscript{$\pm$0.007} & \underline{0.122}\textsubscript{$\pm$0.006} & \textbf{0.115}\textsubscript{$\pm$0.005} \\
    {[Al/Fe]} & \underline{0.053}\textsubscript{$\pm$0.003} & 0.098\textsubscript{$\pm$0.005} & 0.129\textsubscript{$\pm$0.008} & \underline{0.052}\textsubscript{$\pm$0.003} & \underline{0.052}\textsubscript{$\pm$0.003} & 0.100\textsubscript{$\pm$0.004} & \textbf{0.051}\textsubscript{$\pm$0.002} \\
    {[K/Fe]} & 0.159\textsubscript{$\pm$0.008} & \underline{0.137}\textsubscript{$\pm$0.006} & 0.179\textsubscript{$\pm$0.009} & \textbf{0.132}\textsubscript{$\pm$0.006} & 0.157\textsubscript{$\pm$0.007} & \underline{0.138}\textsubscript{$\pm$0.006} & \underline{0.134}\textsubscript{$\pm$0.005} \\
    \addlinespace[3pt]
    {[V/Fe]} & 0.218\textsubscript{$\pm$0.011} & \underline{0.158}\textsubscript{$\pm$0.008} & 0.214\textsubscript{$\pm$0.009} & \underline{0.154}\textsubscript{$\pm$0.009} & 0.207\textsubscript{$\pm$0.009} & \textbf{0.153}\textsubscript{$\pm$0.009} & \textbf{0.153}\textsubscript{$\pm$0.008} \\
    {[Cr/Fe]} & 0.119\textsubscript{$\pm$0.005} & 0.093\textsubscript{$\pm$0.004} & 0.121\textsubscript{$\pm$0.006} & \underline{0.089}\textsubscript{$\pm$0.004} & 0.111\textsubscript{$\pm$0.005} & \underline{0.091}\textsubscript{$\pm$0.004} & \textbf{0.086}\textsubscript{$\pm$0.004} \\
    {[Mn/Fe]} & \underline{0.060}\textsubscript{$\pm$0.002} & 0.074\textsubscript{$\pm$0.003} & 0.086\textsubscript{$\pm$0.005} & \textbf{0.058}\textsubscript{$\pm$0.003} & \underline{0.060}\textsubscript{$\pm$0.003} & 0.072\textsubscript{$\pm$0.003} & \textbf{0.058}\textsubscript{$\pm$0.003} \\
    {[Co/Fe]} & 0.132\textsubscript{$\pm$0.006} & 0.131\textsubscript{$\pm$0.006} & 0.157\textsubscript{$\pm$0.008} & \underline{0.116}\textsubscript{$\pm$0.006} & 0.129\textsubscript{$\pm$0.007} & 0.126\textsubscript{$\pm$0.007} & \textbf{0.115}\textsubscript{$\pm$0.005} \\
    {[Ni/Fe]} & \underline{0.045}\textsubscript{$\pm$0.002} & 0.056\textsubscript{$\pm$0.003} & 0.063\textsubscript{$\pm$0.003} & \underline{0.045}\textsubscript{$\pm$0.002} & \underline{0.046}\textsubscript{$\pm$0.003} & 0.054\textsubscript{$\pm$0.003} & \textbf{0.044}\textsubscript{$\pm$0.002} \\
    \addlinespace[3pt]
    {[Ce/Fe]} & 0.221\textsubscript{$\pm$0.009} & 0.250\textsubscript{$\pm$0.013} & 0.285\textsubscript{$\pm$0.013} & \underline{0.212}\textsubscript{$\pm$0.010} & \underline{0.214}\textsubscript{$\pm$0.010} & 0.235\textsubscript{$\pm$0.012} & \textbf{0.206}\textsubscript{$\pm$0.009} \\
    \bottomrule
  \end{tabular}
\end{table}

\begin{figure}[ht]
  \centering
  \includegraphics[width=\linewidth]{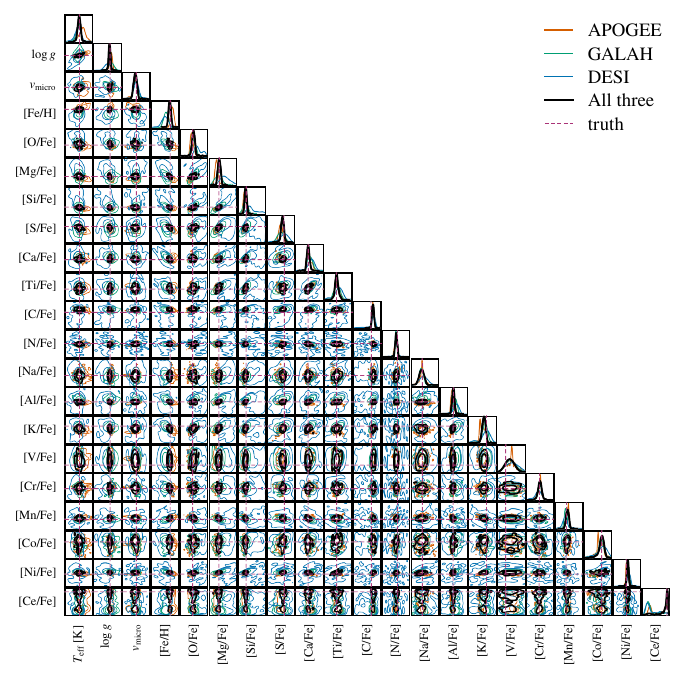}
  \caption{Posteriors for all 21 parameters of the held-out star in Figure~\ref{fig:pairs_heldout}, under each survey alone and all three together. Contours enclose 68\% and 95\%. Dashed lines mark the truth. Best viewed zoomed in.}
  \label{fig:pairs_heldout_all}
\end{figure}

\begin{figure}[ht]
  \centering
  \includegraphics[width=\linewidth]{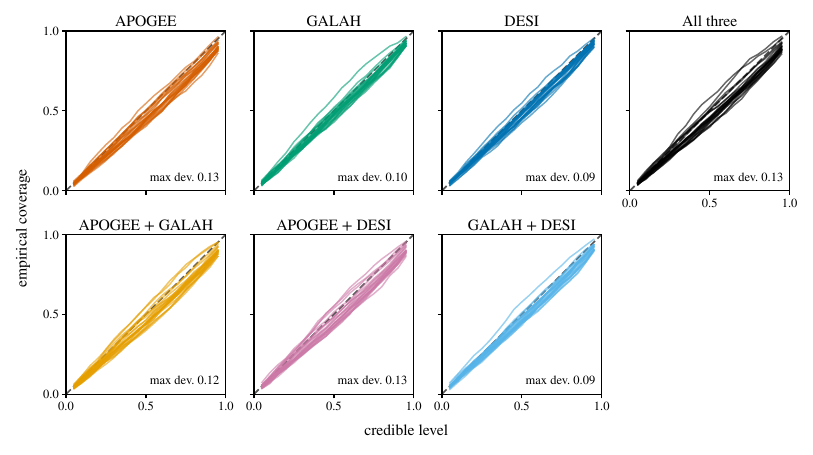}
  \caption{Empirical calibration plot on the held-out simulations. We compare the specified credible interval level (x-axis) against the empirical coverage from our model; lines on one-to-one are calibrated and lines below it are overconfident. We show one line per parameter, and a panel for every survey combination.}
  \label{fig:calibration}
\end{figure}

\end{document}